\documentclass[10pt]{article}
\usepackage{opex2}
\usepackage{times}

\begin{document}
\def\btdplot#1#2{\centering \leavevmode\epsfxsize=#2\textwidth \epsfbox{#1}}
\bibliographystyle{unsrt}

\title{Asymptotic light field in the presence of a
bubble-layer}

\author{Piotr J. Flatau}
\address{Scripps Institution of Oceanography,
University of California, San Diego, La Jolla, California
92093-0221} \email{pflatau@ucsd.edu}

\author{Jacek Piskozub}
\address{Institute of Oceanology,
Polish Academy of Sciences, 81712 Sopot, Poland}
\email{piskozub@iopan.gda.pl}

\author{J. Ronald V. Zaneveld}
\address{Oregon State University,
College of Ocean and Atmospheric Science, Corvallis, OR 97331}
\email{zaneveld@oce.orst.edu}

Optics Express, vol 5, number 5, August 1999,page120-123

Received July 23, 1999; Revised August 26, 1000 (C) 1999 OSA

htpp://epubs.osa.org/oearchive/pdf/11948.pdf \newline

\begin{abstract}
We report that the submerged microbubbles are an efficient source
of diffuse radiance and may contribute to a rapid transition to
the diffuse asymptotic regime. In this asymptotic regime an
average cosine is easily predictable and measurable.
\end{abstract}
\ocis{(010.4450) Ocean optics; (290.4210) Multiple Scattering}

\begin{OEReferences}
\item\label{Zaneveld89a}
J.~R.~V. Zaneveld. ``An asymptotic closure theory for irradiance
in the sea and its inversion to obtain the inherent optical
properties,'' {Limnol. Oceanogr.} {\bf 34},1442--1452 (1989).

\item\label{McCormick95a}
N.~J. Mc{C}ormick. ``Mathematical models for the mean cosine of
irradiance and the diffuse
  attenuation coefficient,''
{Limnol. Oceanogr.} {\bf 40}, 1013--1018 (1995).

\item\label{Bannister92a}
T.~T. Bannister. ``Model of the mean cosine of underwater radiance
and estimation of
  underwater scalar irradiance,''
{Limnol. Oceanogr.} {\bf 37}. 773--780 (1992).

\item\label{Berwald95a}
J.~Berwald, D.~Stramski, C.~D. Mobley, and D.~A. Kiefer.
``Influences of absorption and scattering on vertical changes in
the
  average cosine of the underwater light field,''
{Limnology and Oceanography} {\bf 40}, 1347--1357 (1995).

\item\label{Stramski84a}
D.~Stramski.
\newblock Gas microbubbles: {A}n assessment of their significance to light
  scattering in quiescent seas.
\newblock In {\em Ocean optics {XII} : 13-15 {J}une
  1994, {B}ergen, {N}orway}, Jules~S. Jaffe, editor, Proc. SPIE v. 2258, 704--710 (Bellingham, Wash., USA, 1994).

\item\label{Frouin96a}
R.~Frouin, M.~Schwindling, and P.-Y. Deschamps. ``Spectral
reflectance of sea foam in the visible and near-infrared:
  {I}n situ measurements and remote sensing implications,''
{J. Geophys. Res.} {\bf 101}, 14361--14371 (1996).

\item\label{Mobley94a}
Curtis~D. Mobley.
\newblock {\em Light and water : radiative transfer in natural waters}
\newblock (Academic Press, San Diego, 1994).

\item\label{Flatau98a}
P.~J. Flatau, M.~K. Flatau, J.~R.~V. Zaneveld, and C.~Mobley.
``Remote sensing of clouds of bubbles in seawater,''
\newblock Q. J. Roy. Met. Soc. (1999)(to be published).

\item\label{Farmer84a}
D.~M. Farmer and D.~D. Lemon. ``The influence of bubbles on
ambient noise in the ocean at high wind
  speeds,''
{J. Phys. Oceanogr.} {\bf 14}, 1762--1778 (1984).

\item\label{Thorpe95a}
S.~A. Thorpe. ``Dynamical processes of transfer at the sea
surface,'' {Prog. Oceanogr.} {\bf 35}, 315--352 (1995).

\item\label{Johnson87a}
B.~D. Johnson and P.~J. Wangersky. ``Microbubbles: stabilization
by monolayers of adsorbed particles,'' {J. Geophys. Res.} {\bf
92}, 14641--14647 (1987).

\item\label{Medwin77a}
H.~Medwin. ``In situ acoustic measurements of microbubbles at
sea,'' {J. Geophys. Res.} {\bf 82}, 971--976, (1977).

\item\label{Isao90a}
K.~Isao, S.~Hara, K.~Terauchi, and K.~Kogure. ``Role of
sub-micrometre particles in the ocean,'' {Nature} {\bf 345},
242--244 (1990).

\item\label{Thorpe92a}
S.~A. Thorpe, P.~Bowyer, and D.~K. Woolf. ``Some factors affecting
the size distributions of oceanic bubbles,'' {J. Phys. Oceanogr.}
{\bf 22}, 382--389 (1992).

\item\label{Mulhearn81a}
P.~J. Mulhearn. ``Distribution of microbubbles in coastal
waters,'' {J. Geophys. Res.} {\bf 86}, 6429--6434 (1981).

\item\label{Piskozub94a}
J.~Piskozub.
\newblock Effects of surface waves and sea bottom on self-shading of in-water
  optical instruments.
\newblock In {\em Ocean optics {XII} : 13-15 {J}une
  1994, {B}ergen, {N}orway}, Jules~S. Jaffe, editor, Proc. SPIE v. 2258, 300--308 (Bellingham, Wash., USA, 1994).

\end{OEReferences}

\section{Introduction}
The vertical structure of light in the sea is important in marine
bio-optics. Sunlight is the energy source for the biological food
chain, and the amount and spectrum of solar energy available at a
given depth must be known if accurate productivity calculations
are to be made. The vertical structure of the diffuse attenuation
coefficient in the near surface regime is important in
understanding optical remote sensing. The behavior of the light
field in the sea is described by the equation of radiative
transfer, which relates the light field to the inherent optical
properties of water and its constituents [\ref{Zaneveld89a}]. Once
this equation is solved certain moments of the light field can be
derived and compared with measurements. Of particular interest is
the average cosine of irradiance at depth $z$ which is defined by
$\bar{\mu}(z) = E(z)/E_0(z),$ where $E_0(z)$ is the scalar
irradiance and $E(z)$ is the net (vector) irradiance, both with
units ${\rm Wm^{-2}}$. The downwelling average cosine is
$\bar{\mu}_d(z) = E_d/E_0$, where $E_d$ is the downwelling
irradiance. The average cosine, which can vary between -1 and 1,
gives directional information about the radiance distribution. The
average cosine is also related to the diffuse attenuation
coefficient $K(z)= - {1 / E(z)} {dE(z) / dz}$. Integration of the
radiative transfer equation over all directions leads to ${dE(z) /
dz} = -a(z) E_0(z)$ so that $\bar{\mu}(z)= a(z)/K(z)$, which is
known as Gershun's equation. We can now measure $a(z)$ and $K(z)$
routinely, so that it is possible to verify predictions of the
vertical distribution of $\bar{\mu}$ on the basis of observations.
A number of models of the vertical structure of the average cosine
of the light field have been presented
[\ref{Zaneveld89a}-\ref{Berwald95a}] but they have all been
limited to homogeneous media with a flat surface. On the other
hand, using the data from the Gulf of California and the East
Coast shelf one of us (JRVZ, unpublished) concluded that the
average cosine models do not fit the determination of
$\bar{\mu}(z)$ from $K(z)$ and $a(z)$. It is speculated that this
departure from theory is due to vertical inhomogeneities in the
inherent optical properties (IOPs) and the possibility that the
light field in the surface layer of the ocean may be more diffuse
than theoretical models assume. This diffuse light field may be
due to haze and clouds in the atmosphere, relatively low solar
zenith angles, surface waves, and microbubbles in the surface
layer. A combination of these factors contributes to the light
field being quite diffuse throughout the water column. In this
study we demonstrate the capacity of air bubbles to be an
efficient scatterer.

\section{Results}
There is limited knowledge [\ref{Stramski84a}-\ref{Mobley94a}]
about the radiative transfer properties of bubble clouds, their
inherent optical properties, and their global climatology.
Recently, we reported [\ref{Flatau98a}] on the influence of
submerged bubble clouds within the water on the remote sensing
reflectance. Individual bubble clouds persist for several minutes
and are generated by breaking waves. There is evidence that at
high wind speeds, separate bubble clouds near the surface
coalesce, producing a stratus layer
[\ref{Farmer84a},\ref{Thorpe95a}]. The majority of bubbles
injected into the surface layers of natural waters are unstable,
either dissolving due to enhanced surface tension and hydrostatic
pressures or rising to the air-water interface where they break
[\ref{Johnson87a}]. However, bubbles with long residence times,
i.e. stable microbubbles have been observed
[\ref{Medwin77a}-\ref{Thorpe92a}]. One of the stabilization
mechanisms [\ref{Mulhearn81a},\ref{Johnson87a}] assumes that the
surfactant material is a natural degradation product of
chlorophyll, present in photosynthesizing algae.

\begin{figure}
\btdplot{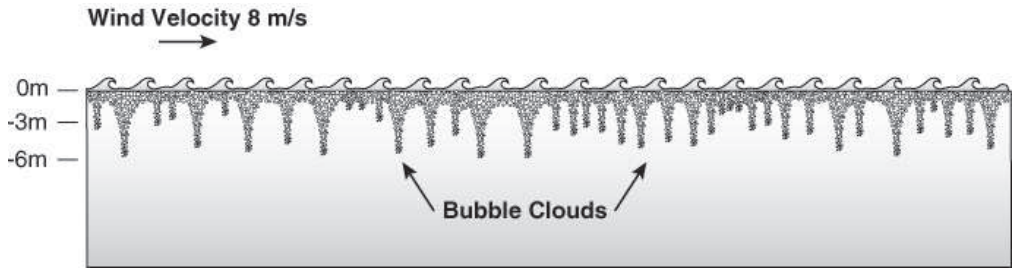}{1.} \caption{\label{bubblesc}
Bubble-stratus layer generated by breaking waves. Bubble
population is continually replenished by wave activity (breaking
waves at the surface) leading eventually to semi-homogeneous
layer.}
\end{figure}

We consider three simple situations: (1) ``infinitely'' deep
homogeneous ocean composed of CDOM, water, and particulates; (2)
``infinitely'' deep homogeneous ocean composed of the same CDOM,
water, and particulates but also with bubbles; (3) two meter layer
of bubbles and ``background'' CDOM, water, and particulates. In
Fig.~\ref{bubblesc} the two-layer case is presented. (based on
[\ref{Thorpe95a}]).

\begin{figure}
\btdplot{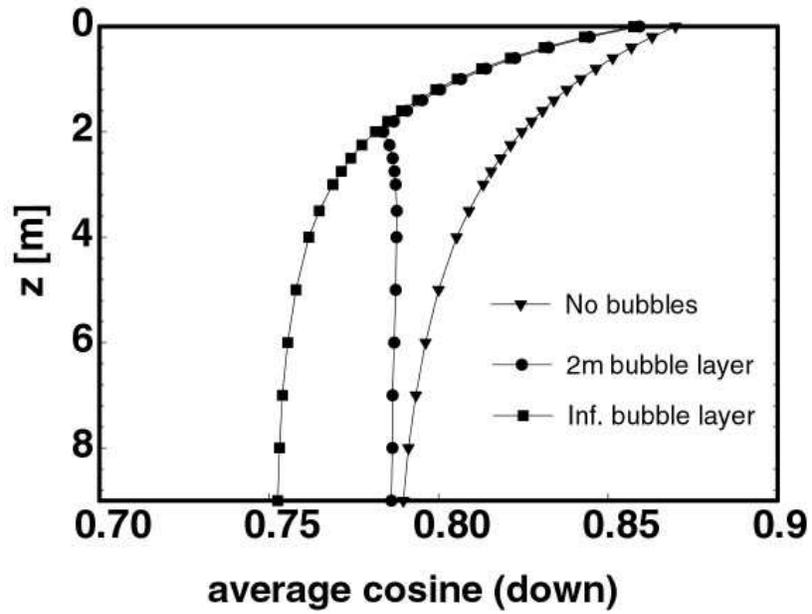}{0.8} \caption{\label{mud} Average cosine
for downwelling radiation $\bar{\mu}_d(z)$ for (1) ``infinitely''
deep homogeneous ocean composed of CDOM, pure water, and
particulates corresponding to chl= $10 {\rm mgm^{-3}}$
(triangles); (2) ``infinitely'' deep homogeneous ocean with
bubbles (cubes); (3) two-layer system composed of background CDOM,
water, and particulates and 2m layer of submerged bubbles close to
the surface (circles). }
\end{figure}

\begin{figure}
\btdplot{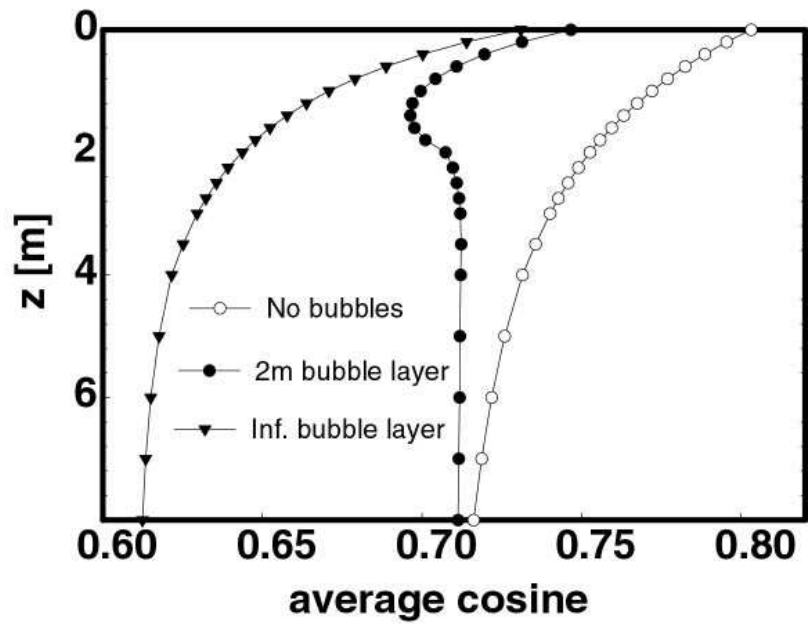}{0.8} \caption{\label{mu} Same as
\protect\ref{mud} but for the average cosine $\bar{\mu}(z)$. }
\end{figure}

The background water, CDOM, and particulates are characterized by
absorption and scattering coefficients $a=0.1801 {\rm m^{-1}}$ and
$b=1.2525 {\rm m^{-1}}$ (thus we include water a and b). It
corresponds to a chlorophyll concentration of $10 {\rm mg m^{-3}}$
at a wavelength $\lambda= 550 {\rm nm}$. The particulate phase
function is that of Petzold for turbid water [7]. The phase
function for bubbles was calculated by averaging from size
parameter $x=10-350$, where $x=2 \pi r / \lambda$, $r$ is radius.
The size distribution follows a $r^{-4}$ law. A phase function was
derived using the exact method for spheres and we assume a
refractive index $n=3/4$. The scattering coefficient for bubbles
is assumed to be $b_{\rm bubble}=0.6181 {\rm m^{-1}}$. In our
recent paper [\ref{Flatau98a}] we discuss in detail the
climatology of bubble layer depth, vertical distribution, and
dependence of depth on the wind speed. Assumptions made here are
typical for $10 {\rm m/s}$ winds in bubble-stratus regime. The
radiative transfer calculations were performed using the
Monte-Carlo technique [\ref{Piskozub94a}]. The main results of
this note are presented in Figs.~\ref{mud}-\ref{mu}.
Fig.~\ref{mud} shows the average cosine for the downwelling
radiation for the three cases discussed above. In this figure
``circles'' are for the two-layer system. At first there a is
rapid decrease in the value of the average cosine as initially
collimated photons are efficiently scattered. After exiting the
``bubble'' layer the light is already diffuse and only a small
adjustment is needed to attain an asymptotic regime for the ``no
bubbles'' situation. This double-exponential transition mechanism
seems to be particularly efficient in establishing the
near-surface diffuse light field. Clearly, the depth of the bubble
layer and the inherent optical properties will further determine
this efficiency. Fig~\ref{mu} shows $\bar{\mu}(z)$, which contains
contribution from the upwelling radiation as well. The upwelling
radiance field adjusts gradually when photons interact with the
``bubble'' layer. There is preferential loss of photons traveling
in the horizontal direction. This leads to more collimated light
and increased average cosine close to the surface. In reference
[8] we discuss in more detail the climatology of bubble clouds,
which strongly depends on wind speed. The importance of bubbles on
an asymptotic light field depends on wind speed, wind gustiness,
and microphysical properties of bubbles. Incoming field projects
and laboratory studies should give a clearer answer on the
influence of bubbles on marine light fields.
In summary we show that the submerged microbubbles are an
efficient source of diffuse radiance and, if present, contribute
to rapid transition to the diffuse asymptotic regime.

\section*{Acknowledgements}
P. J. Flatau was supported in part by the Office of Naval Research
Young Investigator Program. J. Piskozub has been supported in part
by the Institute of Oceanology PAS and by Office of Naval Research
Europe Visiting Scientist Program (VSP). J. Ronald V. Zaneveld
acknowledges support of the Environmental Optics program of the
Office of Naval Research and the Biogeochemistry program of NASA.

\end{document}